\documentstyle[12pt]{article}

\begin{document}
\title{Bound on nonlocal scale from LEP}
\author{N.V.Krasnikov \thanks{E-mail address: KRASNIKO@MS2.INR.AC.RU}
\\Institute for Nuclear Research\\
60-th October Anniversary Prospect 7a,\\ Moscow 117312, Russia}
\date{February 1995}
\maketitle
\begin{abstract}

Leptons, quarks and gauge bosons are assumed to be described by local
field theory. Stringent bound on nonlocal scale can be derived from
the high-precision LEP measurements. We find a bound on nonlocal scale
$\Lambda_{nl} > 1020\: Gev$, C.L. 95\% from LEP data.

\end{abstract}
\newpage

At present the standard explanation of the gauge hierarchy problem is
the supersymmetry with the supersymmetry breaking scale $M_{SUSSY} \leq
O(1)Tev$ \cite{1}. However it is possible to explain the gauge hierarchy
problem by postulation that there is a nonlocal scale $\Lambda_{nl} \leq
O(10)Tev$ \cite{2} in nature.  The main idea of the nonlocal field theory
\cite{3} consists in the use of nonlocal propagator
\begin{equation}
D_{nl}(p^2) = V_{nl}(p^2)(p^2 - m^2 + i\epsilon)^{-1}
\end{equation}
instead of local propagator
$D_{l}(p^2) = (p^2 - m^2 + i\epsilon)^{-1}$. Here $V_{nl}(p^2)$ is an
entire function on $p^2$ decreasing in the euclidean region $p^{2} < 0$
that leads to the ultraviolet convergence of the perturbation theory
in the case of nonlocal field theory \cite{3}. Typical example of
nonlocal formfactor is $V_{nl}(p^2) = \exp(\frac{p^2}{\Lambda_{nl}^{2}})$.
For one-loop correction to the Higgs boson mass we have the following
estimate \cite{2}:
\begin{equation}
\delta m^{2}_{H} = O(\frac{g_{2}^{2}}{16{\pi}^2})\Lambda_{nl}^{2}
\end{equation}
Using the well known upper bound \cite{4} $m_{H} \leq 1 Gev $ on the
Higgs boson mass and requiring that $\delta m^{2}_{H} \leq m^{2}_{H}$
one can find an upper bound for nonlocal scale \cite{2}
\begin{equation}
\Lambda_{nl} \leq O(10)Tev
\end{equation}
So if quantum field theory with nonlocal scale is responsible for
the solution of the gauge hierarchy problem we should expect the existence
of very low nonlocal scale compared to the Planck scale $M_{PL} = 1.2\cdot
10^{19}\: Gev$.

In this note we find a bound on nonlocal scale $\Lambda_{nl} > 1020\: Gev$,
C.L. 95\% from high-precision LEP data. As it has been mentioned before the
main idea of nonlocal field theory consists in the use of nonlocal propagator
(1) instead of local propagator. At the language of Lagrangians nonlocal
propagator (1) corresponds to nonlocal Lagrangian
\begin{equation}
L_{nl} = \frac{1}{2}\phi(x)[V^{-1}(-\Box_{x})(\Box_{x} + m^{2})]\phi(x)
\end{equation}
for scalar particles. Here $\Box_{x} = \partial^{\mu}\partial_{\mu}$.
For the gauge fields the corresponding nonlocal Lagrangian has the form
\begin{equation}
L_{nl,g} = -\frac{1}{2}Tr[F_{\mu\nu}(x)V^{-1}(-\Box_{x})F^{\mu\nu}(x)] \:,
\end{equation}
where
\begin{equation}
F_{\mu\nu(x)} = \partial_{\mu}A_{\nu}(x) - \partial_{\nu}A_{\mu}(x) +
 ig[A_{\mu}(x),A_{\nu}(x)] \: ,
\end{equation}
\begin{equation}
\Box_{x} = (\partial_{\mu} - igA_{\mu}(x))(\partial^{\mu} - igA^{\mu}(x)) \: .
\end{equation}
In this paper we shall assume that nonlocal formfactor $V(p^2)$ is the same
for $SU(3)$, $SU(2)$ and $U(1)$ gauge groups. It appears that in the first
approximation the nonlocale Lagrangian of the type (4) for quarks and leptons
does not lead to the observable effects for the LEP measurements so we neglect
this type of nonlocality. Typical example for the nonlocal formfactor
$V(p^2)$ is
\cite{3}
\begin{equation}
V_{nl}(p^2) = [\frac{\sin(-p^{2}l^{2})^{1/2}}{(-p^{2}l^{2})^{1/2}}]^{2} \: .
\end{equation}
The formfactor (5) has the meaning of the uniformly distributed charge
with the radius l \cite{3}. The introduction of the nonlocal formfactor
$V_{nl}(p^2)$ leads to the propagator for the Z-boson
\begin{equation}
D_{\mu\nu}(p^2) = \frac{g_{\mu\nu} - \frac{p_{\mu}p_{\nu}}{p^2}}
{p^{2}V_{nl}^{-1}(p^2) - M_{Z_{o}}^{2}}
\end{equation}
At present the most accurate and "theoretically clean" LEP observables are
the leptonic widths and leptonic forward-backward asymmetries.
We have the following formulae for the leptonic widths and asymmetries
\cite{5,6}
\begin{equation}
\Gamma_{l} = 2^{0.5}G_{\mu}M_{z}^{3}(g^{2}_{V} + g^{2}_{A})(1 +
\frac{3\alpha}{4\pi})\frac{1}{12\pi} \: ,
\end{equation}
\begin{equation}
A_{FB} = \frac{3g^{2}_{A}g^{2}_{V}}{(g^{2}_{V} + g^{2}_{A})^{2}} \: ,
\end{equation}
where at the tree level
\begin{equation}
g_{A} = -\frac{1}{2} \: ,
\end{equation}
\begin{equation}
g_{V} = -\frac{1}{2} + 2\sin^{2}(\theta_{w}) \: .
\end{equation}
The corresponding one-loop formulae for vector and axial coupling
constants are well known \cite{5,6}. For instance, for the axial
coupling constant in notations of ref.\cite{6} we have
\begin{equation}
g_{A} = (1- \frac{1}{2}\Pi_{w}(0) + \frac{1}{2}\Pi_{Z}(M^{2}_{Z})
-\frac{1}{2}D -\frac{1}{2}\Sigma^{'}_{Z}(M_{Z}^{2}))(-\frac{1}{2}
+2csF_{A}) \; .
\end{equation}
Analogous formula takes place for the vector coupling constant $g_{V}$.
For the nonlocal formfactor $V_{nl}(p^2)$ we have the asymptotic
expansion
\begin{equation}
V_{nl}(p^2) = 1 + \frac{p^{2}}{\Lambda_{nl}} +
O((\frac{p^2}{\Lambda_{nl}^{2}})^{2}) \: .
\end{equation}
One can find using the formula (15) that an account of nonlocal propagator
(9) for Z-boson leads to an additional factor
\begin{equation}
K = 1 + \frac{M^{2}_{Z}}{2\Lambda_{nl}^{2}} + O((\frac{M_{Z}^{2}}
{\Lambda^{2}_{nl}})^{2}) \:
\end{equation}
for axial and vector lepton coupling constants.
We have used the following LEP measurements as input \cite{7}:
the mass of the Z, $M_{Z} = 91.197 \pm 0.007$ Gev, the total width
of the Z, $\Gamma_{Z} = 2.490 \pm 0.007$ Gev, the ratio of the hadronic
partial width to the partial width for Z decay into electron, muon and tau
pairs $R_{e} = 20.76 \pm 0.08$, $R_{\mu} = 20.76 \pm 0.07$, $R_{\tau} =
20.80 \pm 0.08$ and the effective leptonic axial and vector couplings
with Z-boson $g_{A}^{l} = -0.5008 \pm 0.0008$, $g_{V}^{l} = 0.0377 \pm
0.0016$.
In our analysis we used the formulae of ref.\cite{6} to estimate the
standard one-loop corrections. We took the top quark and the Higgs
boson masses in the interval $130  Gev \leq m_{t}\leq 190 Gev$,
$60 Gev \leq m_{H} \leq 500 Gev$. For $\Lambda_{nl}^{2} > 0$ we have found
that
\begin{equation}
\Lambda_{nl} > 1020(1320)\: Gev,\:  C.L.95 \%(C.L.68\%) \:.
\end{equation}
For the less probable from the theoretical point
of view case  $\Lambda_{nl}^{2} < 0 $ (typical formfactor for this case is
$V_{nl}(p^2) = \exp(-\frac{p^2}{\Lambda^2} -
c_{1}(\frac{p^2}{\Lambda^2})^{2})\:$ )
we have found that
\begin{equation}
-\Lambda_{nl}^{2} > (930 Gev)^2((1140Gev)^2),\: C.L.95 \%(C.L.68\%)
\end{equation}
For the nonlocal formfactor(8) bound (17) corresponds to the bound
\begin{equation}
l < 2.4(1.9) \cdot 10^{-17} cm, \: C.L.95 \%(C.L.68\%)
\end{equation}
for nonlocal radius. Note that in ref.\cite{8} bounds on the radii of
quarks and leptons and their weak anomalous magnetic moments have been
derived from the high-precision measurements at LEP and SLC. The authors
of ref.\cite{8} have found bound $R \geq 10^{-17} cm$ for quark and lepton
radii that is very closed to our bound (19).

To conclude, in this note we have found a bound on nonlocal scale
$\Lambda_{nl} > 1020\: Gev$ (C.L.95\%) from LEP measurements.
If the nonlocal scale is responsible for the solution of the gauge
hierarchy problem then it could be discovered at LHC provided that
$\Lambda_{nl} \leq 10 \: Tev$ (the nonlocal glueon propagator will
change the prediction for the two jet cross sections compared to
standard QCD case).

I am indebted to the collaborators of the INR theoretical department
for discussions and critical comments. I thank prof. P.M.Zervas for sending
me preprint \cite{8}. The research described in this publication was made
possible in part by Grant N6G000 from the International Science Foundation
and by Grant 94-02-04474-a of the Russian Scientific Foundation.

\end{document}